\documentclass[conference,letterpaper]{IEEEtran}

\usepackage[utf8]{inputenc} 
\usepackage[T1]{fontenc}
\usepackage{url}
\usepackage{ifthen}
\usepackage{cite}
\usepackage[cmex10]{amsmath} 

\usepackage{amsmath, amssymb, tikz, enumitem, tabularx, bm, mathabx, verbatim, import, subfiles, subfigure, mathtools, tikz-cd, amsthm, stackengine, xcolor}

\newtheorem{proposition}{Proposition}
\newtheorem{lemma}[proposition]{Lemma}

\newtheorem{theorem}[proposition]{Theorem}

\newtheorem{corollary}[proposition]{Corollary}

\theoremstyle{definition}
\newtheorem{definition}[proposition]{Definition}
\newtheorem{example}[proposition]{Example}
\newtheorem{remark}[proposition]{Remark}
\newtheorem{notation}[proposition]{Notation}

\newcommand{\R}{\ensuremath{\mathbb{R}}}
\newcommand{\PR}{\ensuremath{\mathbb{P}}}
\DeclareMathOperator{\Rep}{Rep}
\DeclareMathOperator{\sgn}{sgn}

\usetikzlibrary{intersections}

\begin{document}
\title{Algebraic Representations of Entropy and Fixed-Parity Information Quantities}

\author{
   \IEEEauthorblockN{Keenan J. A. Down\IEEEauthorrefmark{1}\IEEEauthorrefmark{2} and 
                     Pedro A. M. Mediano\IEEEauthorrefmark{3}\IEEEauthorrefmark{4}}
   \IEEEauthorblockA{\IEEEauthorrefmark{1}%
                     Queen Mary University of London}
   \IEEEauthorblockA{\IEEEauthorrefmark{2}%
                     University of Cambridge, kjad2@cam.ac.uk}
   \IEEEauthorblockA{\IEEEauthorrefmark{3}%
                     Imperial College London, p.mediano@imperial.ac.uk}
   \IEEEauthorblockA{\IEEEauthorrefmark{4}%
                     University College London}
}

\maketitle


\begin{abstract}
   Many information-theoretic quantities have corresponding representations in terms of sets. The prevailing signed measure space for characterising entropy, the $I$-measure of Yeung, is occasionally unable to discern between qualitatively distinct systems. In previous work, we presented a refinement of this signed measure space and demonstrated its capability to represent many quantities, which we called logarithmically decomposable quantities. In the present work we demonstrate that this framework has natural algebraic behaviour which can be expressed in terms of ideals (characterised here as upper-sets), and we show that this behaviour allows us to make various counting arguments and characterise many fixed-parity information quantity expressions. As an application, we give an algebraic proof that the only completely synergistic system of three finite variables $X$, $Y$ and $Z = f(X,Y)$ is the XOR gate.
\end{abstract}

\section{Introduction}

The Shannon entropy has many properties which intuitively mirror many set-theoretic identities. The $I$-measure of Yeung, built on earlier work by Hu Kuo Ting, fleshed out the correspondence between expressions of information quantities and set-theoretic expressions via a formal symbolic substitution \cite{yeung1991new, ting1962amount}. Occasionally, however, the $I$-measure can be seen to conflate qualitatively different behaviours. A classic example of two such systems are the dyadic and triadic systems of James and Crutchfield \cite{james2017multivariate}, whose co-information signatures are identical, despite having qualitatively different constructions. In that work, they note that `no standard Shannon-like information measure, and exceedingly few nonstandard methods, can distinguish the two.'

One approach for discerning between these two systems is Partial Information Decomposition (PID), which aims to decompose the mutual information between a series of random source variables $X_1,\ldots, X_n$ and a target variable $T$ into parts \cite{williams2010nonnegative, kolchinsky2022novel, ince2017measuring, bertschinger2014quantifying, rosas2020operational, barrett2015exploration}. These parts are combinations of \textbf{redundant}, \textbf{unique}, and \textbf{synergistic} contributions to $I(X_1,\ldots, X_n; T)$ -- all qualitatively distinct. The last of these pieces, the \textbf{synergistic} information, is information that is provided by multiple sources considered together but no source alone. For example, the perception of depth is not possible unless both eyes are recruited, and hence depth information is conferred \textit{synergistically}. The co-information between three variables is, under the PID framework, the redundant (shared) information minus the synergistic information. For this reason, a negative co-information is a symbol that the system is exhibiting synergistic behaviour.

Multiple versions of the partial information decomposition exist, and no conclusive method of calculating a PID has been selected, mostly likely as such a sensible framework appears to be missing from classical information theory \cite{lizier2018information}. Beloved amongst the practitioners of PID is the XOR gate -- one of the few examples of synergy on which there is nearly unanimous agreement. We will show in this work that the XOR gate is in fact the only system of three variables $X, Y$ and $Z = f(X,Y)$ which can have purely synergistic behaviour.

To do this, we will leverage our previously introduced refined signed measure space $\Delta \Omega$, representing a collection of `atomic' pieces of information \cite{down2023logarithmic, down2024logarithmic}. This decomposition, built from the \textit{entropy loss} when merging variable outcomes into coarser events (see for instance \cite{baez2011characterization} for more context), allowed us to construct a Shannon-like measure which can discern between the Dyadic and Triadic systems of James and Crutchfield \cite{james2017multivariate}. It was also demonstrated that the \textbf{logarithmic atoms} (the components of entropy under this decomposition) have many useful properties.

It was also demonstrated in previous work that making this additional refinement broadens the class of quantities which can be expressed with the measure -- to include, for example, the G\'acs-K\"orner common information \cite{gacs1973common} and the Wyner common information \cite{down2024logarithmic, wyner1975common, xu2013wyner}. We referred to such quantities as \textit{logarithmically decomposable} quantities.

In the present work, we demonstrate that the constructed space $\Delta\Omega$ has much natural algebraic behaviour when considered in tandem with the measure\footnote{In \cite{down2023logarithmic} we used $L^\circ$ to represent this measure, later updating the notation in \cite{down2024logarithmic} to $\mu$, which we use in this work.} $\mu$. We show that the structure of co-information, a standard information quantity \cite{bell2003co}, can be expressed in the form of an order ideal inside of $\Delta \Omega$, and these objects have very stable behaviour under the measure $\mu$.

In addition, we show how the structural properties of $\Delta \Omega$ can be applied to the construction of various entropy inequalities. Throughout we use the OR gate as a motivating example and finish with a result showing that the XOR gate is the only purely synergistic deterministic gate in three variables. That is, given three discrete variables $X$ and $Y$ with $Z = f(X,Y)$ where $X$ and $Y$ are finite, the XOR is the only system with negative co-information for all probability mass functions on $X$ and $Y$.

For a brief recapitulation of the concepts introduced in the previous work \cite{down2024logarithmic}, namely the construction of the signed measure space, we refer the reader to appendix \ref{APPENDIX_Background}. Proofs of all results can be found in appendix \ref{APPENDIX_Proofs}.

\begin{example}
Let $X$ and $Y$ be binary variables with $Z = \text{OR}(X,Y)$.

Recall that the co-information (also known as the \textit{interaction information} \cite{bell2003co, mcgill1954multivariate}) is given by
\begin{align}
I(X;Y;Z) = & H(X)+H(Y)+H(Z) \notag\\
& - H(X,Y) - H(X,Z) - H(Y,Z)\\
& + H(X,Y,Z). \notag
\end{align}
In the case that $P(X = x, Y = y) = 0.25$ for all outcomes $(x,y) \in \Omega$, we have that the co-information is $I(X;Y;Z) \approx -0.19$ bits, being negative in this case. However, in the case that $P(X = 0, Y = 0) = P(X = 1, Y = 1) = 0.45$ and $P(X = 0, Y = 1) = P(X = 1, Y = 0) = 0.05$, then we have $I(X;Y;Z) \approx 0.52$ bits. That is, for this system (and many others), knowledge of the structure of the outcomes alone (that is, any prior knowledge that certain combinations of symbols have zero probability) is not sufficient to determine the sign of the co-information, and it depends upon the underlying probabilities of the system states. We shall see why this is the case for the OR gate in particular in section \ref{SECTION_FixedParity}.
\end{example}

In the section to follow we explore some algebraic perspectives on the structure of $\Delta \Omega$ before exploring how these structures interact with the measure $\mu$ in section \ref{SECTION_IdealMeasure}. Lastly, we apply our results in section \ref{SECTION_FixedParity} to reveal new properties of the XOR gate, demonstrating its uniquely synergistic behaviour.

\section{An Algebraic Perspective on Entropy}
\subsection{Representing information quantities inside $\Delta \Omega$}
\label{SECTION_AlgebraicTheory}

The following definitions for $\Delta \Omega$ and $\Delta X$ are taken from \cite{down2024logarithmic}.

\begin{definition}
\label{DEFINITION_Atom}
Given a discrete outcome space $\Omega$, an \textbf{atom} is a subset $S \subseteq \Omega$ where $|S| \geq 2$. It was shown in \cite{down2024logarithmic} that these atoms represent indivisible contributions to entropy.

The set of all decomposition atoms over $\Omega$ is $\Delta \Omega$.
\end{definition}

For a general set $S$ we use the notation $b_S$ for an atom, but where outcomes are explicitly labelled, e.g. $S = \{1,2,3\}$, we might write $b_{\{1, 2, 3\}}$, $b_{123}$ or simply $123$ where this is clear from context.

It was shown in \cite{down2024logarithmic} that the space $\Delta \Omega$ is a signed measure space under a given measure $\mu$. The exact definition of this measure is given in the appendix.

We now state a key result from our previous work, where we expressed the entropy associated to a random variable $X$ in terms of a subset of $\Delta \Omega$.

\begin{definition}
\label{DEFINITION_Content}
Given a random variable $X$, we define the \textbf{content} $\Delta X$ inside of $\Delta \Omega$ to be set of all atoms of $\Omega$ crossing a boundary in $X$. That is, if $X$ corresponds to a partition $P_1,\ldots, P_n$, then
\begin{multline}
\Delta X = \{ \text{$b_S: S \subseteq \Omega, \exists\,  \omega_i,\omega_j \in S$} \\
\text{with $\omega_i \in P_k$, $\omega_j\in P_l$ such that $k\neq l$ }\}.
\end{multline}
Intuitively, this means that at least two of the outcomes in the atom $b_{\omega_1\ldots\omega_n}$ correspond to distinct events in $X$, although possibly more. We will in general make use of $\Delta$ to represent the logarithmic decomposition functor from random variables to their corresponding sets in $\Delta\Omega$. Note that we often write $123$ to refer to $b_{1, 2, 3}$ for added readability.
\end{definition}

As expected, we have that $\mu(\Delta X) = H(X)$, and we concretise this in a theorem, taken from \cite{down2024logarithmic}.

\begin{theorem}
\label{THEOREM_YeungCorrespondence}
Let $R$ be a region on an $I$-diagram of variables $X_1,\ldots, X_r$ with Yeung's $I$-measure. In particular, $R$ is given by some set-theoretic expression in terms of the set variables $\tilde{X}_1, \ldots, \tilde{X}_r$ under some combination of unions, intersections and set differences.

Making the formal substitution
\begin{equation}
    \tilde{X}_1, \tilde{X}_2, \ldots, \tilde{X}_r \quad \longleftrightarrow \quad \Delta X_1, \Delta X_2, \ldots, \Delta X_r \\
\end{equation}
to obtain an expression $\Delta R$ in terms of the $\Delta X_i$, we have
\begin{equation}
I(R) = \sum_{B \in \Delta R} \mu(B).
\end{equation}
That is, the interior loss measure $\mu$ is consistent with Yeung's $I$-measure.
\end{theorem}

For examples on how this measure can be interpreted geometrically, as well as all proofs of the above results, we refer the interested reader to the previous work \cite{down2024logarithmic}, where we present figures and diagrams demonstrating the geometric and set-theoretic significance of the atoms of our construction.

Many questions about the underlying structure of this space remain to be answered. One peculiarity is that most atoms do not normally appear alone in information quantities. For example, given that an atom $\omega_1 \ldots \omega_n \in \Delta X$ appears in a content, we must also have $\omega_1 \ldots \omega_n \omega_{n+1}\in \Delta X$ appear in the same content, as the definition is just those atoms which, as a set, cross a boundary in $X$. While all atoms have an operational interpretation of crossing boundaries in partitions, individual atoms, at first, do not seem to have much meaning without other atoms in context. Understanding the structural interrelationship between all atoms would allow for a better understanding of structure of different information measures and their relationship to one-another.

In the rest of this section, we explore the structure of our decomposition in the language of posets and \textbf{upper-sets} (or \textbf{ideals}) on those posets, which appear to provide the natural language for the analogous `molecules' to our atoms. We begin by defining an order $\preccurlyeq$ on our atoms, before giving a definition for \textbf{ideals} in $\Delta \Omega$. From there, we will show that all co-informations correspond to ideals and vice versa. We finish this section by characterising the ideals which correspond to the entropy of a variable.

\subsection{Ideals in $\Delta\Omega$}

\begin{definition}
Let $b_{S_1}, b_{S_2} \in \Delta\Omega$ where $S_1, S_2 \subseteq \Omega$. We define a partial ordering on the set $\Delta\Omega$ by setting $b_{S_1} \preccurlyeq b_{S_2}$ whenever $S_1 \subseteq S_2$.
\end{definition}

The following definition is taken from \cite{davey2002introduction}.

\begin{definition}
Given a (partially) ordered set $P$, a subset $J \subseteq P$ is called an \textbf{order ideal}, (\textbf{upper-set}, \textbf{up-set}, \textbf{increasing set}) if, for all $x\in J$ and $y\in P$, we must have $y \in J$ whenever $x \preccurlyeq y$ and $J\neq \varnothing$. That is, $J$ is non-empty and closed under ascending order.

Following standard language, we will say that an ideal $J$ is \textbf{generated} by a collection of elements $g_1,\ldots, g_t$ if for all $b \in J$, we have $g_i \preccurlyeq b$ for at least one $g_i \in \{g_1,\ldots, g_t\}$. We will write $J = \langle g_1,\ldots, g_t \rangle$.

That is to say, the ideal $J$ is the set in $\Delta \Omega$ which contains $g_1,\ldots, g_t$ and all elements which lie above them in the order.
\end{definition}

We note that we deviate from standard nomenclature in this case and refer to these upper-sets simply as `ideals'. In classical order theory, ideals in lattices are down-sets rather than upper-sets and are subject to an additional constraint. In the current work we use \textbf{ideal} in the order-ideal sense, as we expect that future work on $\Delta \Omega$ as a ring might make this definition more intuitive.

We will concern ourselves later with the relationship between the generators of an ideal and the measure of the ideal itself. The following definition for the \textbf{degree} of an atom is taken from \cite{down2024logarithmic}, which we then extend to ideals.

\begin{definition}
Let $b = \omega_1\ldots \omega_d \in \Delta\Omega$. We define the \textbf{degree} of $b$ to be the number of outcomes it contains. That is, $\deg(b) = d$.
\end{definition}

\begin{definition}
We will call $J$ a \textbf{degree $n$ ideal} or \textbf{$n$-ideal} if it can be generated by purely degree $n$ atoms.
\end{definition}

One significant motivation for introducing the language of ideals is to simplify the description of the sets constructed by the decomposition. Rather than writing out the complete set of all atoms, it is often possible to write out the generators of the set as an ideal, vastly reducing the complexity of the notation. Much like ideals in ring theory, it is straightforward to describe the intersection and union of ideals using the generators alone. We introduce some notation and give a proposition to this effect.

\begin{notation}
To further the parallel to ideals in rings, it is sometimes useful to introduce multiplicative notation for the union of two atoms. That is, we will make use of the notation
\begin{equation}
b_{S \cup T} = b_S b_T.
\end{equation}
For example, using the shorthand notation from before, we have $123 \cdot 234 = 1234$.
\end{notation}

\begin{proposition}
\label{PROPOSITION_IdealAlgebra}
Let $G = \{g_1, \ldots, g_n\}$ be a set of generators for the ideal $I = \langle G \rangle = \langle g_1, \ldots, g_n \rangle$, and let $H = \{h_1, \ldots, h_m\}$ be a set of generators for the ideal $J = \langle H \rangle = \langle h_1, \ldots, h_m \rangle$, where $I$ and $J$ are ideals inside $\Delta \Omega$. Then:

\begin{multline}
I \cup J = \langle g_1, \ldots, g_n, h_1, \ldots, h_m \rangle \\
= \langle\, b \mid b \in G \cup H \,\rangle,
\end{multline}

\begin{multline}
I \cap J = \langle g_1 h_1, g_1 h_2, \ldots, g_n h_{m-1}, g_n h_m \rangle \\
= \langle\, gh \mid g \in G, h \in H \,\rangle
\end{multline}

This formulation mimics the natural behavior of ideals in rings.
\end{proposition}

\begin{remark}
It is occasionally convenient for notation to consider ideals generated by single outcomes $\omega$ even though we formerly excluded these and $\varnothing$ from $\Delta \Omega$. We may alternate between including and excluding these atoms for algebraic simplicity. Recall that the singlet and empty atoms do not contribute to the entropy, so this choice of notation does not affect the measure.\footnote{We note also that including these entities would endow $\Delta \Omega$ with the complete structure of a lattice, which, although currently not required, might be useful for future work.}
\end{remark}

\begin{example}
Let $\Omega = \{1, 2, 3, 4\}$. Then $\Delta\Omega = \{12, 13, 14, 23, 24, 34, 123, 124, 134, 234, 1234\}$. Inside of $\Delta \Omega$, an ideal consists of all atoms which `contain' a generator. For example, the ideal $\langle 12, 13\rangle = \{12, 13, 123, 124, 134, 1234\}$ is a $2$-atom ideal, as it is generated by degree 2 atoms.
\end{example}

\subsection{Representation of quantities with ideals}

We should justify that these ideals are a natural object of study. As it turns out, all entropy expressions without multiplicity and without conditioning are given by ideals, which follows from the next lemma.

\begin{lemma}
\label{LEMMA_co-informationContentIsIdeal}
Given any co-information $I(X_1; \ldots; X_t)$ (including entropy and mutual information), the corresponding content $\Delta X_1 \cap \cdots 
 \cap \Delta X_t$ is an ideal.
\end{lemma}

\begin{example}
It is worth noting that ideals themselves do not, in general, have corresponding partitions, but every partition has a corresponding ideal. As an example for how to conceptualise these `sub-partitions,' consider the system $\Omega = \{1, 2, 3\}$ where $X$ has partition $\{\{1\}, \{2, 3\}\}$ and $Y$ has partition $\{\{1,3\}, \{2\}\}$ as per figure \ref{FIGURE_TrianglePartition}.

We have that
\begin{align}
\Delta X &= \{12, 13, 123\} \text{ and} \\
\Delta Y &= \{12, 23, 123\}.
\end{align}

Taking the mutual information between these sets corresponds algebraically to the intersection $\Delta X \cap \Delta Y = \langle 12 \rangle = \{12, 123\}$.

We note that this upper-set $\langle 12 \rangle$ corresponds to the ability to discern between 1 and 2, but not between 1 and 3, or 2 and 3. Moreover, the upper-set $\langle 12\rangle$, despite not representing a partition itself, gives the mutual information when measured, i.e. $\mu(\langle 12 \rangle) = I(X;Y)$.

That is to say, generalising from the language of partitions to the language of ideals has allowed us to properly describe mutual information -- a quantity which partitions cannot in general represent.

\begin{figure}[!t]
    \centering
        \newcommand{\figmult}{0.5}
        \newcommand{\distance}{7}
    
        \begin{tikzpicture}
        \node at (-\distance,1*\figmult) {$3$};
        \node at (-\distance - 0.8660254*\figmult, -0.5*\figmult) {$1$};
        \node at (-\distance+0.8660254*\figmult, -0.5*\figmult) {$2$};
        \draw (-\distance,1)--(-\distance+0.8660254, -0.5)--(-\distance-0.8660254, -0.5)--cycle;
        \draw (-\distance,0)--(-\distance, -0.5);
        \draw (-\distance,0)--(-\distance - 0.8660254/2, 0.25);
        \draw (-\distance,0)--(-\distance + 0.8660254/2, 0.25);

        \node at (-\distance*0.6666,1*\figmult) {$3$};
        \node at (-\distance*0.6666 - 0.8660254*\figmult, -0.5*\figmult) {$1$};
        \node at (-\distance*0.6666+0.8660254*\figmult, -0.5*\figmult) {$2$};
        \draw (-\distance*0.6666,1)--(-\distance*0.6666+0.8660254, -0.5)--(-\distance*0.6666-0.8660254, -0.5)--cycle;
        \draw (-\distance*0.6666,0)--(-\distance*0.6666, -0.5);
        \draw (-\distance*0.6666,0)--(-\distance*0.6666 - 0.8660254/2, 0.25);

        \node at (-\distance*0.3333,1*\figmult) {$3$};
        \node at (-\distance*0.3333 - 0.8660254*\figmult, -0.5*\figmult) {$1$};
        \node at (-\distance*0.3333+0.8660254*\figmult, -0.5*\figmult) {$2$};
        \draw (-\distance*0.3333,1)--(-\distance*0.3333+0.8660254, -0.5)--(-\distance*0.3333-0.8660254, -0.5)--cycle;
        \draw (-\distance*0.3333,0)--(-\distance*0.3333, -0.5);
        \draw (-\distance*0.3333,0)--(-\distance*0.3333 + 0.8660254/2, 0.25);

        \node at (0,1*\figmult) {$3$};
        \node at (-0.8660254*\figmult, -0.5*\figmult) {$1$};
        \node at (0.8660254*\figmult, -0.5*\figmult) {$2$};
        \draw (0,1)--(0.8660254, -0.5)--(-0.8660254, -0.5)--cycle;
        \draw (0,0)--(0, -0.5);

        \draw[dotted] (-0.83333*\distance, -0.5)--(-0.83333*\distance, 1) ;

        \node at (-0.5*\distance, 0.2) {$\cap$};
        \node at (-0.3333/2*\distance, 0.2) {$=$};       

        \node at (-0.6666*\distance, 1.4) {$X$};
        \node at (-0.3333*\distance, 1.4) {$Y$};
        \node at (-\distance, 1.4) {$\Omega$};
        
        \end{tikzpicture}
        
        \caption{An outcome space $\Omega = \{1,2,3\}$ and two variables $X$ and $Y$ defined over $\Omega$. In this case, the intersection of the contents $\Delta X \cap \Delta Y$ is given by the ideal $\langle 12\rangle$. That is to say, $I(X;Y) = \mu(\langle 12 \rangle)$. If the mutual information could be represented by a partition in this case, we would get something like the above intersection. This is, of course, impossible in the language of partitions, but valid in ideals.}
        \label{FIGURE_TrianglePartition}
\end{figure}
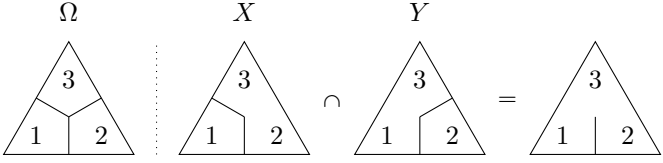

\end{example}

As it turns out, the converse to lemma \ref{LEMMA_co-informationContentIsIdeal} also holds, which we state here.

\begin{theorem}
\label{THEOREM_co-informationExpressions}
Let $\Omega$ be a finite outcome space and let $\{X_a: a\in A\}$ be the collection of all possible random variables defined on $\Omega$ (indexed by $A$). Then there is a one-to-one correspondence
\begin{equation}
\{\text{ideals in $\Delta \Omega$}\} \leftrightarrow \{\text{possible co-informations on $\Omega$}\}
\end{equation}
where the co-informations are given by $I(X_1;\ldots; X_j)$ for any number of arbitrary variables $X_1,\ldots, X_j$ defined on $\Omega$.
\end{theorem}

This result tells us that for any valid co-information on some collection of variables defined on an outcome space $\Omega$, then there is a corresponding ideal in $\Delta\Omega$, and for every ideal in $\Delta\Omega$, there is a corresponding collection of variables which give the resulting co-information.  As an immediate side-effect of this result, we have an alternative derivation of proposition 33 in \cite{down2024logarithmic}. By reasoning carefully with ideals, we are able to make a stronger statement about subsets of $\Delta \Omega$.

\begin{corollary}
\label{COROLLARY_EntropyExpressions}
Let $\Omega$ be a finite outcome space. Then there is a one-to-one correspondence
\begin{equation}
\{\text{subsets of $\Delta\Omega$}\} \leftrightarrow \left\{\text{\stackanchor[4pt]{entropy expressions}{without multiplicity on $\Omega$}}\right\},
\end{equation}
where by an \textit{`entropy expression without multiplicity'} we mean an expression of the form
\begin{equation}
\sum_{\text{$P$ partitioning $\Omega$}} n_P H(P)
\end{equation}
for $n_P \in \mathbb{Z}$ where no region is double-counted in any $I$-diagram.
\end{corollary}

Note that for the purposes of these two results, we do not consider the singlet atoms $\{\omega\}$ or the empty set to be elements of $\Delta \Omega$, as they contribute no entropy (see \cite{down2024logarithmic} for more justification).

This result shows that, with clever inclusion and exclusion, it is always possible to extract individual atoms as classical entropy expressions on variables in $\Delta\Omega$. That is, they form a natural basis for entropy expressions. As such, the atoms of $\Delta \Omega$ would be uniquely placed for a module-theoretic or vector-space perspective on information.

Since these atoms appear to be a natural basis for entropy expressions, if we count them without multiplicity, we are able to determine how many expressions for information exist without accounting for the same contribution multiple times. We give a corollary to this end.

\begin{corollary}
\label{COROLLARY_CountingEntropyExpressions}
Given a finite outcome space $\Omega$ with $|\Omega| = n$, there are $2^{2^n - n - 1}$ possible classical entropy expressions without multiplicity.
\end{corollary}

Counting with multiplicity, we can see that the space of all entropy expressions on $\Omega$ is a free module over $\mathbb{Z}$, where the atoms $b$ form a very natural basis.

We now state a practical result which tells us, intuitively, exactly which ideals correspond to partitions and how we can find the generators of the ideal corresponding to a finite random variable $X$. For the purpose of this result, it is again useful to consider the singlets $\{\omega\}$, but the resulting representation of $\Delta X$ will not contain them.

\begin{theorem}
\label{THEOREM_VariableIdealRepresentation}
Let $X$ be a discrete random variable on the outcome space $\Omega$, where $X$ has corresponding partition $Q_t: t \in T$ for some indexing set $T$ of parts $Q_t$. Then $\Delta X$ as an ideal is given by
\begin{equation}
\Delta X = \bigcup_{\substack{a,b\, \in \,T \\ a\,\neq \, b}} \langle \{\omega \in Q_a\} \rangle \cap \langle \{ \omega \in Q_b \} \rangle.
\end{equation}
We note in particular that in posets, the union of order ideals is equal to the order ideal with the union of their generators. Equivalently we have
\begin{multline}
\Delta X = \bigcup_{t\,\in\, T} \,\langle \{\omega \in Q_t\} \rangle \cap \langle \{ \omega \in Q_t^c \} \rangle \\
= \bigcup_{t \in T} \,\langle \{\omega \bar{\omega}: \omega \in Q_t, \bar{\omega}\in Q_t^c \}\rangle
\end{multline}
In particular, $\Delta X$ as an ideal is generated by $2$-atoms.
\end{theorem}

\begin{example}
Consider the outcome space $\Omega = \{1, 2, 3, 4\}$. Now let $X$ be the variable with partition $\{\{1, 2\}, \{3\}, \{4\}\}$. Then as an ideal we have
\begin{align}
\Delta X &= (\langle 1, 2 \rangle \cap \langle 3 \rangle) \cup (\langle 1, 2 \rangle \cap \langle 4 \rangle) \cup (\langle 3 \rangle \cap \langle 4 \rangle)\notag\\
&= \langle 13, 23 \rangle \cup \langle 14, 24 \rangle \cup \langle 34 \rangle \\
&= \langle 13, 23, 14, 24, 34 \rangle \notag
\end{align}
\end{example}

\begin{corollary}
\label{COROLLARY_Degree2GivesStructure}
Let $X_a, a\in A$ be a family of discrete variables on the outcome space $\Omega$. Knowledge of how the 2-atoms are located among the $\Delta X_a$ is sufficient to describe how all other atoms are located.
\end{corollary}

Restating this, knowledge of the 2-atoms contained in each $\Delta X_a$ is sufficient to deduce the presence of any atom in any set-theoretic expression constructed using the $\Delta X_a$.

We have now successfully described the structure of entropy through the algebraic lens of ideals in a poset and illustrated that ideals in this lattice correspond to co-informations, while other subsets of $\Delta \Omega$ correspond to entropy expressions on $\Omega$.

To illuminate the power of this flavour of the theory, in the next section we shall see how these ideals interact with the measure $\mu$ and use our results to demonstrate that mutual information is always given by a degree 2 ideal. Not only this, but we will give a generalisation which bounds the degree of the generators for ideals representing the intersection of more than 2 variables. We then extend these techniques to explore ideals giving fixed-sign information quantities. This intriguing result will show that a surprising amount can be learned about an information quantity without much knowledge of the underlying probabilities.

\section{Properties of the Measure on Ideals}
\label{SECTION_IdealMeasure}

We have now developed lots of language for discussing the ideals inside of the lattice $\Delta \Omega$. Moreover, having seen that co-information is perfectly described by these ideals, it would be a natural question to ask how the measure $\mu$ interacts with the ideal structure. In this section we will demonstrate that the entropy contribution of an ideal can, much like atoms, be neatly categorised as either positive or negative in many cases, and we shall see that this provides various tools for constructing new bounds.

In this section we begin by demonstrating that the mutual information is always given by a degree 2 ideal. To do this, we shall need the following notion of \textit{restriction}, which we shall utilise in the proofs to follow.

\begin{definition}
Let $X$ be a random variable on a finite outcome space $\Omega$, and let $S \subseteq \Omega$. We define the \textbf{restriction}\footnote{In \cite{down2024logarithmic} we used the notation $\cdot|_S$ for restriction. Here we use $\cdot_S$ to simplify some of the notation.} of a collection of atoms $W \subseteq \Delta \Omega$ to $S$ by
\begin{equation}
W_S = \{b_Q \in W: Q\subseteq S\}.
\end{equation}

In particular, we will use the notation $\Delta X_S$ and $\langle \ldots \rangle_S$ to construct contents and ideals inside of restrictions.
\end{definition}

Restriction simply allows us to focus our attention on a subset of the atoms -- in particular, those whose outcomes all belong to the restricting subset $S$. Note that, given some subset $W \subseteq \Delta \Omega$, we have that $W_S \subseteq W$, rather than operating with an entirely new class of atoms.

One of the strengths of the measure $\mu$ above that of entropy $H$ alone is that $\mu$ is homogeneous and works across multiple scales. As such, every statement and piece of structure given here for $\Delta \Omega$ and ideals therein \textit{also} applies to $\Delta \Omega_S$ for $S \subseteq \Omega$. We shall demonstrate that many problems exploring the intersection of ideals (and hence the intersection of entropies) can be much simplified by restricting. We proceed with the first result on mutual information, where we use this concept in the proof.

\begin{theorem}[Mutual information is a degree 2 ideal]
\label{THEOREM_MutualInformationDegree2}
Let $X$ and $Y$ be two random variables. Then there exists a set of 2-atom generators $\{a_ib_i : a_i, b_i \in \Omega \}$ for $i=1,\ldots, k$ such that 

\begin{equation}
I(X;Y) = \mu( \langle a_1 b_1, \ldots, a_kb_k \rangle )
\end{equation}
\end{theorem}

We have demonstrated something rather intriguing; mutual information looks a lot like a normal variable content in that it is always generated by degree 2 atoms, but the generators of mutual information do not need to correspond to a representable subset of $\Delta \Omega$. When working with ideals in general, one would expect that the intersection between generators of degree $m$ and $n$ would have degree bounded above by $m+n$, so it is rather surprising that the mutual information has this property.

Extending the investigation of the G\'acs-K\"orner common information in \cite{down2024logarithmic}, it can now be seen that:

\begin{corollary}
\label{COROLLARY_CommonInfoIsIntersectionOfDegree2}
The G\'acs-K\"orner common information is generated by degree 2 atoms, and the generating set is the largest subset of generators of the mutual information which is representable.\footnote{\label{FOOTENOTE_Representable}\textbf{Representable} meaning a subset of $\Delta \Omega$ which can be represented by a random variable (or partition) on $\Omega$. In the original work \cite{down2023logarithmic} we referred to this property as \textbf{discernibility}.}
\end{corollary}

This result confirms our natural intuition for selecting generators to construct a variable. We provide also a generalisation of theorem \ref{THEOREM_MutualInformationDegree2} to co-information.

\begin{theorem}
\label{THEOREM_GeneratorDegreeBound}
Let $\Omega$ be the joint outcome space of $M$ discrete variables $X_1,\ldots, X_M$. Then the content of $I(X_1;\ldots; X_M)$ can be completely generated by atoms of degree at most $M$.
\end{theorem}

This result states that the degree of the generators of an ideal corresponding to some co-information is always bounded above by the number of variables. This result vastly reduces the search space of generators when studying the properties of co-information, and we make use of it in our study of fixed-parity systems in the next section.

\begin{example}
Consider the standard OR gate given by outcomes $(X,Y,Z=\text{OR}(X,Y))$, which we label as follows:

\begin{center}
\begin{tabular}{| c | c | c | c |}
\hline
$X$ & $Y$ & $Z = \text{OR}(X,Y)$ & Outcome $(\omega)$ \\
\hline
0 & 0 & 0 & 1 \\
0 & 1 & 1 & 2 \\
1 & 0 & 1 & 3 \\
1 & 1 & 1 & 4 \\
\hline
\end{tabular}
\end{center}

\begin{figure}[ht]
\begin{center}

\begin{tikzpicture}
    \newcommand{\circrad}{2.5cm}
    \newcommand{\scaleshift}{1}
    
    \begin{scope}
        \clip (-0.866*\scaleshift,0.5*\scaleshift) circle (\circrad);
        \clip (0.866*\scaleshift,0.5*\scaleshift) circle (\circrad);
        \fill[lightgray!30!white] (0*\scaleshift,-1*\scaleshift) circle (\circrad);
    \end{scope}

    \draw (-0.866*\scaleshift,0.5*\scaleshift) circle (\circrad) node at (-2.8*\scaleshift,2.8*\scaleshift) {$X$};
    \draw (0.866*\scaleshift,0.5*\scaleshift) circle (\circrad) node at (2.8*\scaleshift,2.8*\scaleshift){$Y$};
    \draw (0*\scaleshift,-1*\scaleshift) circle (\circrad) node at (0*\scaleshift,-3.9*\scaleshift) {$Z$};

    \node at (-1.8,-1) {13};
    \node at (1.8,-1) {12};
    
    \node at (0,2.4) {23};
    \node at (0,1.9) {234};
    
    \node at (0,-0.4) {14};
    \node at (0.5,0.2) {134};
    \node at (-0.5,0.2) {124};
    \node at (0,0.8) {1234};
    
    \node at (0,-1) {123};
    
    \node at (2.7*0.866,2.7*0.5) {34};
    \node at (-2.7*0.866,2.7*0.5) {24};
    
\end{tikzpicture}

\caption{An $I$-diagram demonstrating the entropy structure for the OR gate. The shaded region corresponds to the ideal $\langle 14, 123\rangle$. Note that in this case the degree of the generators is bounded above by 3, as we have the intersection of 3 variables as per theorem \ref{THEOREM_GeneratorDegreeBound}.}
\label{FIGURE_ORStructure}
\end{center}
\end{figure}
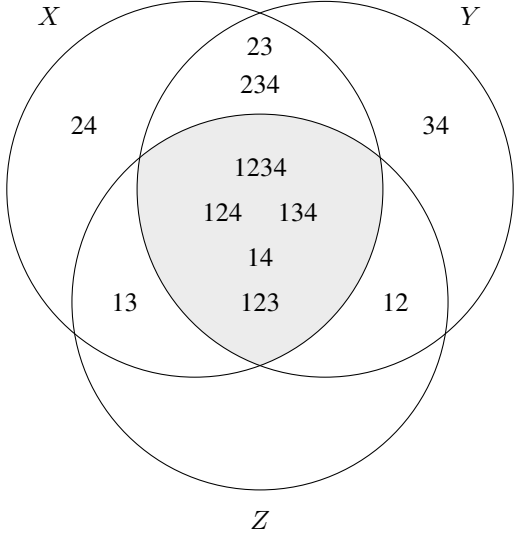

Note that in this instance we have
\begin{equation}
I(X;Y;Z) = \mu(\langle 14, 123 \rangle)
\end{equation}
which is generated by at most degree 3 atoms, as expected. Note that we are discussing the structure of the OR gate without any mention of the probabilities. A diagram representing the structure is given in figure \ref{FIGURE_ORStructure}. As expected, the degree of the generators is bounded above by 3.
\end{example}

Although this is an interesting representation of the structure of the co-information between random variables, we haven't said much yet about the relationship between these ideals and their measures $\mu(J)$. As it turns out, for certain classes of ideals, the sign of $\mu(J)$ is just as easy to characterise as the signs of the atoms themselves.

\begin{lemma}
\label{LEMMA_SingletIdealSign}
Let $J = \langle \omega_1\cdots \omega_d\rangle$ be an ideal generated by a single degree $d$ atom with $\PR(\omega_1),\ldots, \PR(\omega_d) \neq 0$. Then $(-1)^{d}\mu(J) > 0.$
\end{lemma}

This result is quite powerful as it tells us that, in certain scenarios, we can know the sign of the ideal and the information measure it represents without any knowledge of the probabilities. We will strengthen this result shortly to demonstrate that certain classes of ideals, which we call strongly fixed-parity ideals, have fixed-parity measures.

\begin{definition}
Let $J = \langle g_1,\ldots, g_j\rangle$ be an ideal.  If $j = 1$, we say that $J$ is \textbf{strongly fixed-parity}, and set the \textbf{parity} of $J$ as $P(J)=(-1)^{\deg g_1}.$

Moreover, if $j\geq 2$, we shall say that $J$ has \textbf{strongly fixed-parity} if there is an expression
\begin{equation}
\mu(J) = P(J) \sum_{\alpha \in A} P(J_\alpha)\mu(J_\alpha)
\end{equation}
for some finite collection of fixed-parity ideals $\{J_\alpha: \alpha \in A\}$, with the equality holding across all probability distributions $\mathbb{P}$ on $\Omega$ and some $P(J) \in \{-1, 1\}$, which we call the \textbf{parity} of $J$.

Lastly, we shall say that an ideal $J \subseteq \Delta\Omega$ is of \textbf{strongly mixed-parity} if it has generators of both even and odd degree.
\end{definition}

\begin{example}
The ideal $\langle 12, 23\rangle$ is strongly fixed even parity, as
\begin{align}
\mu(\langle 12, 23 \rangle) &= \mu(\langle 12 \rangle) + \mu(\langle 23 \rangle) - \mu(\langle 12 \rangle \cap \langle 23 \rangle) \notag\\
&= \mu(\langle 12 \rangle) + \mu(\langle 23 \rangle) - \mu(\langle 123 \rangle).
\end{align}
The ideal $\langle 123 \rangle$ has strong negative (odd) parity, and the two degree 2 ideals have strong positive (even) parity. When composing the parities, the measure of the whole set is given by three positive parts, so it makes sense to call $\langle 12, 23\rangle$ positive fixed-parity.
\end{example}

\begin{theorem}[Ideals of Strong Parity]
\label{THEOREM_IdealParity}
Let $J$ be an ideal in $\Delta \Omega$. If $J$ is of strongly even parity, then $\mu (J) \geq 0$, and if $J$ is of strongly odd parity, then $\mu (J) \leq 0$.
\end{theorem}

This result is most pleasant as it reflects what feels like a natural intuition for how these systems should behave. In particular, given a system of variables $X_1,\ldots, X_n$, defined by partitions on a finite outcome space $\Omega$, information quantities which reflect strongly fixed-parity ideals have a predetermined sign, for any underlying probability distribution over $\Omega$.

In this section we demonstrated that the algebraic construction of ideals in our poset from the previous section plays remarkably well with the measure $\mu$ of our construction, and we developed several tricks for manipulating expressions in $\Delta \Omega$. We build on this theory in the next section and apply it to the problem of finding purely synergistic systems of the form $X$, $Y$, and $Z = f(X,Y)$.

\section{Fixed-Parity Systems}
\label{SECTION_FixedParity}

We begin with a definition to connect the algebraic perspective to information quantities.

\begin{definition}
Let $X_1,\ldots, X_n$ be a system of variables on a finite outcome space $\Omega$. We say that the system $X_1,\ldots, X_n$ is a \textbf{fixed-parity}\footnote{We use \textbf{fixed-parity} and \textbf{mixed-parity} in this paper to refer to the sign of the co-information. We use the terms \textbf{strongly fixed-parity} and \textbf{strongly mixed-parity} to refer to an algebraic property of ideals. Our objective is to show that the algebraic property implies the sign property.} system if the sign of
\begin{equation}
I(X_1;\ldots; X_n)
\end{equation}
is fixed, regardless of the underlying probability distribution. Similarly, we say that an entropy expression $E(X_1,\ldots, X_n)$ defined on $\Omega$ has \textbf{fixed-parity} if its sign is always fixed, regardless of the underlying probabilities in $X_1,\ldots, X_n$. We say that the system is \textbf{negative/odd fixed-parity} if the co-information is always negative, and \textbf{positive/even fixed-parity} if the co-information is always positive.
\end{definition}

There is a natural dual question to be asked here: is it possible to have a fixed-parity system where the co-information is a strongly-mixed ideal -- that is, generated by elements of varying degrees? The next theorem shows this is, in fact, impossible.

\begin{theorem}
\label{THEOREM_FixedParityCoiIffFixedParityIdeal}
A system of variables $X_1,\ldots, X_n$ with $I(X_1; \ldots; X_n)$ given by an ideal of strongly mixed-parity cannot have a fixed parity.

That is to say, being strongly-mixed (an algebraic property) implies mixed sign (a property of the measure on the set).
\end{theorem}

This theorem gives a partial converse to theorem \ref{THEOREM_IdealParity}, in that it gives us a way to characterise some mixed-parity systems. Although we have not characterised every fixed-parity or mixed-parity system, we expect such a characterisation in terms of ideal properties (algebraic reasoning alone) should be possible. However, the tools we have already constructed are sufficient to show our final result.

\begin{theorem}
\label{THEOREM_FixedParityXOR}
The only negative fixed-parity (always synergy-dominated) system given by two finite variables $X, Y$ and a deterministic function $Z = f(X,Y)$ is the XOR gate.
\end{theorem}

\section{Conclusion}

In this paper we extended our results on the measure $\mu$ from \cite{down2023logarithmic, down2024logarithmic} to an algebraic construction inside of the space $\Delta \Omega$. We demonstrated that in many cases the study of `ideals' (in the order-theoretic sense) inside of $\Delta \Omega$ simplifies bounding problems, and we showed that these ideals form a natural intermediate language between partitions and have useful behaviour in tandem with the measure $\mu$.

While in the present work the issue of bounding is applied to the study of fixed-parity quantities, we expect these techniques can be used in multiple scenarios where bounding over all possible probabilities is required. Moreover, we expect that there is a stronger version of theorem \ref{THEOREM_IdealParity} which might be stated with weaker restrictions on the underlying ideals. Future work may develop this theory.

Lastly, we applied our results to show that the XOR gate is the only purely synergistic system (i.e. always possessing negative co-information) of finite variables $X, Y$ and $Z = f(X,Y)$ for a deterministic function $f$ (see also \cite{jansma2023higher}). In particular, we did this with an algebraic proof that did not require any navigation of the space of probability distributions.

We hope this work might be applied to the problem of Partial Information Decomposition (PID) \cite{mediano2021towards, rosas2020operational, kolchinsky2022novel, ince2017measuring, james2017multivariate} and contribute to the widening body of knowledge in set-theoretic information theory.

\subsubsection*{Acknowledgements} We would like to thank Dan Bor and Abel Jansma for helpful input on the contents of this work and future applications.

\bibliographystyle{plain}
\bibliography{main}

\clearpage
\begin{appendices}

\section{Background on the measure}
\label{APPENDIX_Background}

In the previous work we introduced the signed measure space $(\Delta \Omega, \mu)$\footnote{Note that in the original work \cite{down2023logarithmic} we used $L^\circ$ to represent our measure. In \cite{down2024logarithmic} we opted to use $\mu$ for greater clarity, which we use in this and future work.}, where the sigma algebra is taken implicitly as the set of all subsets of $\Delta \Omega$. To start, we restate the definition of the space $\Delta \Omega$ as was given in the first paper \cite{down2023logarithmic}.

\begin{definition}
Let $(\Omega, \mathcal{F}, P)$ be a finite probability space where the sigma-algebra $\mathcal{F}$ is given by all subsets of $\Omega$. Then we define the \textbf{complex} of $\Omega$ (or \textit{content}) to be the simplicial complex on all outcomes $\omega \in \Omega$, with the vertices removed:
\begin{equation}
\Delta\Omega = \bigcup_{k = 2}^N \Omega_k \cong \mathcal{P}(\Omega) \setminus \left( \{\{\omega\}: \omega \in \Omega\} \cup \{\varnothing\} \right)
\end{equation}
\end{definition}

This space contains $2^{|\Omega|}-|\Omega|$ elements (called \textbf{atoms}) for a given finite outcome space $\Omega$.

In order to construct the signed measure space $(\Delta\Omega, \mu)$ we must also define the measure. In the original work this representation of the measure is given as a proposition. We give it here as the primary definition.

\begin{definition}
\label{DEFINITION_InteriorLoss}
Let $T = \{p_1,\ldots, p_k\}$ for some collection of probabilities corresponding to outcomes $\omega_1,\ldots, \omega_k$, and hence corresponding to the atom $\{\omega_1,\ldots, \omega_k\}$. For clarity we write
\begin{equation}
\sigma(T) = \sigma(p_1,\ldots, p_k) = (p_1 + \cdots + p_k)^{(p_1 + \cdots + p_k)}.
\end{equation}
To further simplify we write
\begin{equation}
A_k = \prod_{\substack{S \subseteq\{p_1,\ldots, p_n\} \\ |S| = k}} \sigma(S).
\end{equation}
Then the measure is given by
\begin{equation}
\label{ClearerExpression}
\mu(p_1,\ldots, p_n) = \sum_{k=1}^n (-1)^{n-k}\log (A_k)
\end{equation}
\end{definition}
This definition arises from the perspective of entropy loss, which has appeared previously in the literature and has some natural advantages over the classical formulation of entropy \cite{baez2011characterization}. The measure given here is constructed using two steps; firstly, by considering the entropy loss $L$ when a number of outcomes $\omega_1,\ldots, \omega_t$ are merged, so that they are not distinguishable by any system, and secondly, by performing a M\"obius inversion on this partially ordered set of subsets of $\Omega$, ordered under inclusion.

The purpose of the M\"obius inversion is not transparently clear. However, using the measure of loss $L$ alone, while useable to derive a measure space previously seen by Campbell in \cite{campbell1965entropy}, does not possess sufficient resolution to capture all information quantities one naturally wishes to capture, such as the mutual information and co-information. Incorporating the M\"obius inversion breaks the construction into smaller pieces which multiple systems might share, despite not being extractable using the loss alone.

The loss function, as given immediately by a result of Baez et al. \cite{baez2011characterization}, is homogeneous of degree $d$ when applied to the $d$-th Tsallis entropy \cite{tsallis1988possible}. By extension, the measure $\mu$, which can be viewed as an alternating sum of the losses $L$, is also homogeneous of degree $d$ when built on the $d$-th order Tsallis entropy. Moreover, the measure $\mu$ has some intriguing properties, which we shall briefly restate here. The interested reader should refer to the original publication \cite{down2023logarithmic} for more detail.

\begin{lemma}
\label{LEMMA_InteriorLossAt0}
For $p_1,\ldots, p_n, x \in \R^+$ where $n\geq 0$, we have
\begin{equation}
\lim_{x\to 0} \mu(p_1,\ldots, p_n, x) = 0
\end{equation}
\end{lemma}

This lemma guarantees that the measure disappears and becomes null if any of the constituent probabilities are zero.

\begin{lemma}
\label{LEMMA_MuAtInfinity}
Let $p_1,\ldots, p_{n-1}, x \in \R^+$ and let $x$ vary. Then
\begin{equation}
\lim_{x \to \infty} | \mu(p_1,\ldots, p_{n-1}, x)| = |\mu(p_1,\ldots, p_{n-1})|
\end{equation}
\end{lemma}

This result showed that if one of the `probabilities' tends to infinity, then the size of the entropy contribution tends towards that of an atom lying \textit{beneath it}.

Lastly, as a particularly intriguing property of the measure, its sign is known on all atoms of the partial order.

\begin{theorem}
\label{THEOREM_AlternatingDerivatives}
Let $p_2,\ldots, p_{n} \in \R^+$ be a sequence of nonzero arguments for $n\geq 2$ and $m\geq 0$. Then
\begin{equation}
(-1)^{m+n} \frac{\partial^m \mu}{\partial x^m}(x, p_2,\ldots, p_n) \geq 0.
\end{equation}
\end{theorem}

Setting $m = 0$ it becomes clear that the sign of the measure $\mu$ on a given atom $\omega_1, \ldots, \omega_n$ is dependent only on the number of outcomes $n$. The co-information, by contrast, is not a fixed-parity quantity in general. For example, given three random variables, the co-information can be positive, negative, or even dependent on the underlying probabilities.

Coupled with lemma \ref{LEMMA_MuAtInfinity}, we have that $\mu$ varies monotonically between 0 and the magnitude of the atoms beneath it.

\begin{corollary}[Magnitude can only decrease]
\label{COROLLARY_MuMagnitude}
Let $p_1,\ldots, p_{n-1}, \tau \in \R^+\cup\{0\}$ for $n\geq 3$. Then
\begin{equation}
|\mu(p_1,\ldots, p_{n-1}, \tau)| < |\mu(p_1,\ldots, p_{n-1})|
\end{equation}
\end{corollary}

This corollary is intriguing in that it bounds the contribution to the entropy of an atom by all of the atoms which lie under it in the partial order. This can be thought of as the notion that `higher order contributions to the entropy are bounded above by lower order contributions to the entropy.'

\section{Proofs for Results}
\label{APPENDIX_Proofs}

The proof for theorem \ref{THEOREM_YeungCorrespondence} and results \ref{LEMMA_InteriorLossAt0} to \ref{COROLLARY_MuMagnitude} can be found in \cite{down2024logarithmic}, where we also give an alternative expression for definition \ref{DEFINITION_InteriorLoss}.

\subsection*{Proof of proposition \ref{PROPOSITION_IdealAlgebra}}
\begin{proof}
Suppose $b \in I \cup J$. Then either $b \succcurlyeq g$ or $b \succcurlyeq h$ for some $g \in G$ or $h \in H$. Hence $b \in \langle g_1,\ldots, g_n, h_1, \ldots, h_m \rangle$ as needed. Conversely if $b \in \langle g_1,\ldots, g_n, h_1,\ldots, h_m \rangle$, then $b \succcurlyeq b'$ for some $b' \in \{g_1,\ldots, g_n, h_1, \ldots, h_m\}$, so $b$ must be contained in either $I$ or $J$, so $b \in I\cup J$.

Suppose that $b \in I$ and $b \in J$. Then there exists $g \in G$ and $h\in H$ with $b \succcurlyeq g$ and $b \succcurlyeq h$. Hence $b \succcurlyeq gh$, so $b \in \langle g_1h_1,\ldots, g_n h_m\rangle$. Conversely, if $b \in \langle g_1h_1,\ldots, g_nh_m \rangle$ then there exists a generator $gh$ of $\langle g_1h_1,\ldots, g_nh_m\rangle$ with $b \succcurlyeq gh$. Since $b\succcurlyeq gh$, we must have $b\succcurlyeq g$ and $b\succcurlyeq h$, so $h \in I$ and $h \in J$, so $h \in I\cap J$.
\end{proof}

\subsection*{Proof of Lemma \ref{LEMMA_co-informationContentIsIdeal}}
\begin{proof}
By the definition of content, we must have that $\Delta X_i$ is an ideal. Moreover, the intersection of ideals is an ideal, so the co-information $I(X_1;\ldots; X_t)$ must also correspond to an ideal.
\end{proof}

\subsection*{Proof of Theorem \ref{THEOREM_co-informationExpressions}}
Firstly we note that all co-informations correspond to ideals by the previous result in lemma \ref{LEMMA_co-informationContentIsIdeal}. It suffices to show that all ideals in $\Delta \Omega$ correspond to some co-information. As such, given an ideal $J$ we need to find a collection of variables $X_1,\ldots, X_t$ where $I(X_1;\ldots; X_t) = \mu(J)$.

Ideals are unique up to their generators, and we only need to consider sets of generators which are not contained in each other, otherwise one of them is not needed as a generator. For each generator $g_i$, let $n_i = \deg g_i$. Let $S_i \subseteq \Omega$ be the set of elements in $g_i$. Then consider the $2^{n_i}-2$ variables given by the partitions
\begin{equation}
X_{ij} = \{ (\Omega \setminus S_i) \cup Q_{ij} , S_i \setminus Q_{ij} \}
\end{equation}
for every $Q_{ij} \subset S_i$ a non-empty proper subset of $S_i$, with $j \in \{1, \ldots, 2^{n_i} - 2\}$. Intuitively these variables spread the elements in $g_i$ across two partitions in every combination possible, so that the only guaranteed boundary crosses consistent across the entire collection $X_{ij}$ are by the $g_i$ atom and atoms in $\langle g_i\rangle$. Then
\begin{equation}
\bigcap_{\varnothing \, \subset \, Q_{ij} \, \subset \, S_i} \Delta X_{ij} = \langle g_i \rangle.
\end{equation}
Here we write $Q_{ij}$ to symbolise that these partitions are taken to obtain atom $g_i$. Across the set of generators $g_1,\ldots, g_k$, we consider all possible products
\begin{equation}
Y_{j_1 j_2 \cdots j_k} = X_{1j_1} \wedge X_{2j_2} \wedge \cdots \wedge X_{kj_k} = \bigwedge_{1 \leq i \leq k} X_{ij_i}.
\end{equation}
Where any combination from the $j_i$ where $1\leq j_i \leq 2^{n_i}-2$ can be taken. Note that we write $A \wedge B$ to mean the coarsest partition which is finer than $A$ and $B$. In practice, every variable corresponds to choosing one of the $Q_j$ for each generator, so every single combination is represented as a variable. Then
\begin{equation}
\bigcap_{j_1,\ldots, j_k} \Delta X_{j_1j_2 \ldots j_k} = \langle g_1,\ldots, g_k\rangle
\end{equation}
exactly, as any other generators will be removed, giving the result.

\subsection*{Proof of Corollary \ref{COROLLARY_EntropyExpressions}}
\begin{proof}
Given any atom $b$ we can consider the ideal $I = \langle b \rangle$ and then using conditioning (in the sense of a set difference of information) we may subtract the higher co-information $C = \{\bigcup_{\omega \neq b} \langle b\omega \rangle\}$, which is itself an ideal as the unions of ideals in lattices are ideals. We have that $C \subseteq I$, meaning we may condition it out in order to obtain the expression $I \setminus C = \{b\}$. That is, $b$ alone populates some region on an $I$-diagram between all variables $X_a$.

Taking any collection of atoms hence corresponds to a collection of regions on the maximal $I$-diagram, provided they are not counted with multiplicity.
\end{proof}

\subsection*{Proof of Corollary \ref{COROLLARY_CountingEntropyExpressions}}
\begin{proof}
There are $2^n - n - 1$ elements in $\Delta \Omega$, as the points and the empty set do not contribute to the entropy, leaving $2^n - n - 1$ atoms, and hence $2^{2^n-n-1}$ possible entropic expressions without multiplicity, including the zero expression.
\end{proof}

\subsection*{Proof of Theorem \ref{THEOREM_VariableIdealRepresentation}}
\begin{proof}
Precisely those atoms $\Delta X$ are all those atoms which cross a boundary in $X$, that is, they must contain the pair $\omega_a \omega_b$ for $\omega_a \in P_a$ and $\omega_b \in P_b$ where $P_a$ and $P_b$ are different parts in the partition. The atom $\omega_a \omega_b$ can be written as the intersection of these two prime ideals.

Since the union of ideals $I_1$ and $I_2$ is the ideal generated by the union of their generators in lattices, we have that $\Delta X$ must be the union across these parts and hence generated by 2-atoms.

The second expression follows quickly from the first; every atom in $Q_a^c$ must lie in some other part $Q_b$ and vice versa.
\end{proof}

\subsection*{Proof of Corollary \ref{COROLLARY_Degree2GivesStructure}}
\begin{proof}
All other atoms are described by whether or not they are contained in the intersection of the variable ideals $\Delta X_a$, which by the previous result are generated by degree 2 atoms. Hence the knowledge of how these are distributed will describe the distribution of all other atoms.
\end{proof}

\subsection*{Proof of Theorem \ref{THEOREM_MutualInformationDegree2}}
\begin{proof}
We know that $\Delta X$ and $\Delta Y$ are both degree 2 ideals as they are given by union of intersections of prime degree 1 ideals, so their intersection $\Delta X \cap \Delta Y$ can have generators of at most degree 4. Hence we need to demonstrate that for every degree 3 or degree 4 generator in $\Delta X \cap \Delta Y$ there is a degree 2 generator which contains it.

We demonstrate that every degree 4 atom in $\Delta X \cap \Delta Y$ is contained in a degree 2 ideal. The argument for the degree 3 atoms is very straightforward and uses the same trick. Suppose that we have a degree 4 atom $\omega_1 \omega_2 \omega_3 \omega_4$ which crosses a boundary in $X$ and in $Y$. We may restrict to just these four outcomes $\omega_1, \omega_2, \omega_3, \omega_4$, on which the partition of $X$ and the partition of $Y$ must now also restrict to a partition.

Since $\omega_1\omega_2 \omega_3 \omega_4$ is contained in $\Delta X_{\{1,2,3,4\}}$ and $\Delta Y_{\{1,2,3,4\}}$, we must have that the local partition of $X$ and the local partition of $Y$ are non-trivial, so that $\omega_1\omega_2\omega_3\omega_4$ crosses a boundary in this partition.

Without loss of generality, the potential local partitions of any random variable $\Delta Q_{\{1,2,3,4\}}$ up to reordering of the $\omega_i$ are given by
\begin{align}
&\langle 12, 13, 14 \rangle \notag\\
&\langle 13, 23, 14, 24 \rangle \notag\\
&\langle 12, 13, 14, 23, 24 \rangle \notag\\
&\langle 12, 13, 14, 23, 24, 34 \rangle
\end{align}

In particular, the total number of possible degree-2 generators in 4 outcomes is $^4C_2 = 6$, so taking the intersection of $\Delta X_{\{1,2,3,4\}} \cap \Delta Y_{\{1,2,3,4\}}$ will, by the pigeonhole principle, have a degree 2 atom in their intersection unless both $\Delta X_{\{1,2,3,4\}}$ and $\Delta Y_{\{1,2,3,4\}}$ have at most 3 degree 2-atoms. Of the four possibilities above, only the first satisfies this possibility, so both $X$ and $Y$ are of this form.

Without loss of generality, we assume $\Delta X_{\{1,2,3,4\}}$ is given by $\langle 12, 13, 14\rangle$. The only possible degree 2 ideal which does not intersect with $\langle 12, 13, 14\rangle$ is given by $\langle 23, 24, 34 \rangle$, so we should expect that $\Delta Y_{1, 2, 3, 4} = \langle 23, 24, 34\rangle$. However, this does not correspond to any partition on $\{1,2,3,4\}$, as it does not contain a generator containing element 1. Thus $Y$ cannot have this form and we must have that $\Delta X_{\{1,2,3,4\}} \cap \Delta Y_{\{1,2,3,4\}}$ must intersect and contain a degree 2 element, so the degree 4 atom $\omega_1\omega_2\omega_3\omega_4$ is contained in a degree 2 ideal.

For any degree 3 atom the argument is even simpler; the smallest possible ideal $\Delta X_{\{1,2,3\}}$ must have either 2 or 0 generators of degree-2 when restricted. If it had 0 generators, then $\{1,2,3\}$ cannot cross a boundary in $X$, so would not be present in $\Delta X \cap \Delta Y$. Hence we must have $\Delta X_{\{1,2,3\}}$ has at least 2 generators of degree 2, with the same being true for $Y$. As such, they must intersect with each other at a degree-2 atom by the pigeonhole principle, as the total number of possible generators is $^3C_2 = 3$.
\end{proof}

\subsection*{Proof of Corollary \ref{COROLLARY_CommonInfoIsIntersectionOfDegree2}}
\begin{proof}
Using a result from \cite{down2023logarithmic}, we have that $C_{\mathrm{GK}}(X; Y) = \Rep(\Delta X \cap \Delta Y)$ (the maximally representable subset inside of $\Delta X\cap \Delta Y$ - see footnote \ref{FOOTENOTE_Representable}). We have now also shown that both $\Delta X \cap \Delta Y$ and $\Rep(\Delta X \cap \Delta Y)$ are degree 2 ideals. Hence the generators of the representable subset must be a subset of the generators of the mutual information.
\end{proof}

\subsection*{Proof of Theorem \ref{THEOREM_GeneratorDegreeBound}}
\begin{proof}
We proceed by induction on $M$, by showing that the theoretical minimum number of generators must still be large enough to force an overlap. We have demonstrated in the previous theorem that the statement is true for $M=2$. Suppose that the statement is true for $M-1$, then the ideal corresponding to the co-information $I(X_1; \ldots; X_{M-1})$ for the first $M-1$ variables has generators of degree at most $M-1$. Multiplying the generators of $\Delta I(X_1; \ldots; X_{M-1})$ by the generators for $\Delta X_M$, we hence know that $\Delta I(X_1;\ldots; X_M)$ can be generated by atoms of at most degree $M+1$. Hence we need to show that any degree $M+1$ atom is actually contained in a degree $M$ ideal.

We will use a similar counting argument to result \ref{THEOREM_MutualInformationDegree2}. In particular, given a finite set of size $k$ and two subsets of size $a_1$ and $a_2$, the minimum size of their intersection is given by $a_1+a_2-k$. Given three subsets, a minimum size for the intersection is then given by $(a_1+a_2-k) + a_3 - k$, and so on. Hence given $l$ subsets the corresponding expression is 

\begin{equation}
\label{EQUATION_MinimumIntersection}
a_1 + \ldots + a_{l} - k(l-1).
\end{equation}

Suppose that $\omega_1\ldots \omega_{M+1}$ is a degree $M+1$ atom contained in the co-information  $\Delta I(X_1,\ldots, X_M)$, which we need to demonstrate is contained in a degree $M$ ideal. Restricting to $\{\omega_1,\ldots, \omega_{M+1}\}$, the minimum number of degree $M$ atoms in $\Delta X_{i, \{1, \ldots, M+1\}}$ for any $i$ is given when $X_i$ corresponds locally to a partition of the form $\{\{\omega_i\}, \{\omega_i\}^c\}$ for some single $\omega_i \in \Omega$ [If this isn't immediately clear, consider any partition of $\Omega$ - we could choose a coarser sub-partition into 2 parts which must contain fewer 2-atoms, so minimising the number of 2-atoms overall is equivalent to finding the minimum number of degree 2 atoms in a partition of $\Omega$ into 2 parts. This is equivalent to minimising the value of $k\cdot(|\Omega|-k) = k|\Omega| - k^2$ for $0 < k < |\Omega|$, which happens at $k = 1$ or $k = |\Omega| - 1$].

Hence there must be a minimum of $^MC_{M-1} = M$ degree $M$ atoms in $\Delta X_{i, \{1, \ldots, M+1\}}$ (as we have already selected one outcome from the $M+1$ available outcomes - now we must select the other $M-1$ outcomes). The maximum size of the set of all possible degree $M$ atoms in the restriction to $\{1,\ldots, M+1\}$ is $^{M+1}C_M = M+1$.

Hence taking the intersection of $M$ variables, assuming a minimal number of degree $M$ atoms, and using the expression in equation \eqref{EQUATION_MinimumIntersection}, we need to only to demonstrate that
\begin{equation}
M \cdot M - (M-1) \cdot (M+1) > 0,
\end{equation}
which always evaluates to unity, proving that there is at least one degree $M$ ideal containing every degree $M+1$ atom in the intersection, proving the result.
\end{proof}

\subsection*{Proof of Theorem \ref{LEMMA_SingletIdealSign}}
\begin{proof}
Let our ideal be $\langle \omega_1\ldots \omega_k\rangle$. We will proceed by induction on the difference $d = |\Omega| - k$, arguing at each step that the upper-set is monotonic in probability of the last element $\omega_{k+d}$. We note that the sign of the upper-set might only change if it contains additional outcomes, so provided we treat this carefully, we can also allow $\Omega$ to vary (via restriction), provided that it always contains $\omega_1,\ldots, \omega_k$.

As earlier, we will write $\langle \omega_1 \ldots \omega_k \rangle_S$ to illustrate that we are operating inside some restricting set $S$. These quasi-ideals are quite justified, as all of the previous results must still hold even if we assume the probabilities inside of $S$ do not sum to 1. These atoms will still have the same measure, regardless of the context $S$ in which we find them.

For the first case $|\Omega| = k$, we note that $\langle \omega_1\ldots \omega_k \rangle_{\{\omega_1,\ldots, \omega_k\}}$ consists of the single atom $\omega_1\ldots \omega_k$. We will use the shorthand notation $\langle \omega_1,\ldots, \omega_k\rangle_{\{\omega_1,\ldots, n\}} = \langle \omega_1,\ldots, \omega_k \rangle_n$ for some simplicity. By theorem \ref{THEOREM_AlternatingDerivatives}, which characterises the sign of individual atoms, we have both that $(-1)^k \mu(\langle \omega_1\ldots \omega_k\rangle_k) >0$ for $\PR(\omega_k)\neq 0$, and that $\mu(\langle \omega_1\ldots \omega_k \rangle_k)$ varies monotonically in $\PR(\omega_k)$ between 0 and $-\mu(\omega_1,\ldots, \omega_{k-1})$. So the theorem is true for $d = 0$.

Now suppose that $\mu(\langle \omega_1\ldots \omega_k\rangle_{k+d})$ varies monotonically in $\PR(\omega_{k+d})$ between 0 and $-\mu(\langle \omega_1\ldots \omega_k\rangle_{k+d-1})$. Then we first note that

\begin{align}
\begin{split}
\label{t_expression}
\langle \omega_1\ldots \omega_k\rangle_{\{\omega_1,\ldots, \omega_{k+d}, \omega_{k+d+1}\}} = \langle \omega_1 \ldots \omega_k \rangle_{\{\omega_1,\ldots, \omega_{k+d}\}} \\
\cup \,\, \omega_{k+d+1}\, \langle \omega_1 \ldots \omega_k \rangle_{\{\omega_1,\ldots, \omega_{k+d}\}}.
\end{split}
\end{align}

Where we use the multiplicative notation to signify that $\omega_{k+d+1}$ is added as an outcome to all atoms in $\langle \omega_1 \ldots \omega_k\rangle$. For example:

\begin{equation}
4 \cdot \langle 12 \rangle_{\{1,2,3\}} = \{124, 1234\}.
\end{equation}

Hence we can view $\mu(\langle\omega_1\ldots \omega_{k}\rangle)$ as a function on $\PR(\omega_{k+d+1}) = p_{k+d+1}$:

\begin{align}
\begin{split}
\mu(\langle \omega_1\ldots \omega_k\rangle_{\{\omega_1,\ldots, \omega_{k+d+1}\}}) = \mu(\langle \omega_1\ldots \omega_k \rangle_{\{\omega_1,\ldots, \omega_{k+d}\}}) \\
+ \mu(\omega_{k+d+1} \langle \omega_1 \ldots \omega_k \rangle_{\{\omega_1,\ldots, \omega_{k+d}\}})
\end{split}
\end{align}

We now notice that the second term can be expressed

\begin{align}
\begin{split}
\label{equation_ideal_as_function_of_last}
\mu(\omega_{k+d+1} \langle \omega_1 \ldots \omega_k \rangle_{\{\omega_1,\ldots, \omega_{k+d}\}}) \\
= \mu(\langle \omega_1 \ldots \omega_k \omega_{k+d+1} \rangle_{\{\omega_1,\ldots, \omega_k, \omega_{k+d+1}\}})
\end{split}
\end{align}

But now we can see that the difference between $k+1$ and $k+d+1$ is just $d$, so this reduces to the case for $d$. By assumption, we hence have that this ideal varies monotonically on $\omega_{k+d+1}$ between $0$ and $\mu(\langle \omega_1 \ldots \omega_k\rangle_{k+d} )$.

This means that the entire expression in equation \eqref{equation_ideal_as_function_of_last} must monotonically vary between 0 and $\mu(\langle \omega_1 \cdots \omega_k \rangle)$ as a function of $\omega_{k+d+1}$.

Since we can construct each ideal by successively increasing $d$ and this leaves the sign intact (note that no probability tends to infinity), the sign is left unchanged, proving the result.
\end{proof}

\subsection*{Proof of Theorem \ref{THEOREM_IdealParity}}
\begin{proof}
Let $J = \langle g_1,\ldots, g_j\rangle$ be an ideal of strong even/positive parity, with the result for odd/negative parity following equivalently. By definition of strong even parity, we must have
\begin{equation}
\mu(J) = \sum_{\alpha \in A} P(J_\alpha) \mu(J_\alpha)
\end{equation}
Every strong fixed-parity ideal is defined in terms of an a finite sum of ideals with one generator, so we may assume without loss of generality that the $J_\alpha$ have single generators.

By virtue of lemma \ref{LEMMA_SingletIdealSign}, we know that $P(J_\alpha) = \sgn(\mu(J_\alpha))$. Hence, taking the sum across all the $J_\alpha$, we have that all of the terms $P(J_\alpha)\mu(J_\alpha)$ must be positive. As all terms in the sum positive, so too is $\mu(J)$.
\end{proof}

\subsection*{Proof of Theorem \ref{THEOREM_FixedParityCoiIffFixedParityIdeal}}
\begin{proof}
We will first allow ourselves to consider probabilities not summing to one, demonstrating that the sign has a given parity, and then we shall scale appropriately using the homogeneity property of $\mu$ to obtain meaningful probabilities once more, while the parity shall be fixed.

Suppose $J$ is a strongly mixed ideal. Then $J$ has an even degree generator $g$. We first send all atoms in $\Omega \setminus g$ (as a set) to 0. Then we have
\begin{equation}
\mu \langle J \rangle = \mu \langle g \rangle > 0.
\end{equation}
Summing the probabilities, we let $K = \sum_{\omega \in g} \PR(\omega)$. Then we scale
\begin{equation}
0 \leq \frac{1}{K}\mu(\{\PR(\omega): \omega \in g\}) = \mu\left( \left\{ \frac{\PR(\omega)}{K}: \omega \in g\right\}\right)
\end{equation}
Where we now have $\sum_{\omega \in g} \frac{\PR(\omega)}{K} = 1$. Hence we have found a given set of probabilities where $\mu(J) > 0$.

Repeating the exercise for $g$ of odd degree will similarly show that there are probabilities such that $\mu(J) < 0$. Hence $\mu(J)$ can be either positive or negative given a strongly mixed-parity ideal, giving the result.
\end{proof}

\subsection*{Proof of Theorem \ref{THEOREM_FixedParityXOR}}
\begin{proof}
We begin by briefly demonstrating that $Z=XOR(X,Y)$ has co-information $\Delta I(X;Y;Z)$ given by a strongly odd parity ideal. Given the outcomes \begin{center}
\begin{tabular}{|c|c|c|c|}
\hline
$X$ & $Y$ & $Z$ & $\omega$ \\
\hline
0 & 0 & 0 & 1 \\
0 & 1 & 1 & 2 \\
1 & 0 & 1 & 3 \\
1 & 1 & 0 & 4 \\
\hline
\end{tabular}
\end{center}
we have that $\Delta X \cap \Delta Y \cap \Delta Z = \langle 123, 124, 134, 234 \rangle$. In this case we have
\begin{align}
\mu(\langle 123, 124, 134, 234 \rangle) & = \mu(\langle 123, 124 \rangle) + \mu(\langle 134, 234 \rangle) \notag\\
& \quad - \mu(\langle 123, 124 \rangle \cap \langle 134, 234 \rangle) \notag\\
& = \mu(\langle 123, 124 \rangle) + \mu(\langle 134, 234 \rangle) \notag\\
& \quad - \mu(\langle 1234 \rangle)
\end{align}
where now we also have
\begin{align}
\mu(\langle 123, 124 \rangle) & = \mu(\langle 123 \rangle) + \mu(\langle 124 \rangle) - \mu(\langle 1234 \rangle) \\
\mu(\langle 123, 234 \rangle) & = \mu(\langle 123 \rangle) + \mu(\langle 234 \rangle) - \mu( \langle 1234 \rangle).
\end{align}
Working backwards, we see that $\langle 123, 124\rangle$ and $\langle 134, 234 \rangle$ are negative (odd) fixed-parity ideals, so that $\Delta X \cap \Delta Y \cap \Delta Z$ in this case is a negative fixed-parity ideal.

To show that there are no other such deterministic functions on three variables, we start by considering the case where $X$ and $Y$ are both binary variables. In this case, we know that we can express all events on $Z$ in terms of $f(X,Y)$ on the four outcomes
\begin{center}
\begin{tabular}{|c|c|c|}
\hline
$X$ & $Y$ & $\omega$ \\
\hline
0 & 0 & 1 \\
0 & 1 & 2 \\
1 & 0 & 3 \\
1 & 1 & 4 \\
\hline
\end{tabular}
\end{center}
In this case, we have $\Delta X \cap \Delta Y = \langle 14, 23\rangle$, which is known to have positive measure as it reflects a mutual information. Similarly, any subset $\langle 14 \rangle$ or $\langle 23 \rangle$ will also have positive measure, so we cannot have an ideal generated by a degree 2 ideal alone. However, the ideal cannot have even and odd generators (as then it would have mixed parity by theorem \ref{THEOREM_FixedParityCoiIffFixedParityIdeal} and it cannot have generators more than degree 3 by theorem \ref{THEOREM_GeneratorDegreeBound}. Hence the ideal must be exclusively generated by degree 3 atoms and we must nullify these two degree-2 atoms.

Hence we know that in order to have degree 3 atoms generating $I(X;Y;Z)$, $Z$ must have equal values on the outcome pairs $\{1,4\}$ and $\{2,3\}$. Moreover, we cannot have that $f(0,0) = f(0,1) = f(1,0) = f(1,1)$ for all outcomes, as then $I(X;Y;Z) = 0$. Hence we must have that $f(0,0) = f(1,1) = 0$ and $f(0,1) = f(1,0) = 1$, that is, $Z$ is the XOR gate.

We now extend by induction to give the full result. Let $N_X$ be the number of events in $X$ and $N_Y$ the total number of events in $Y$. In the case where either $N_X$ or $N_Y$ is 1, then that variable must be constant and have zero entropy, so the co-information $I(X;Y;Z)$ is trivially zero.

We have seen that in the case where $N_X = N_Y = 2$ that the only negative fixed-parity system of the form $X, Y, f(X,Y)$ is the XOR gate. Without loss of generality we consider the case $N_X = 3$ and $N_Y = 2$ without loss of generality to highlight the inductive argument. In this case again we know that $Z$ can be computed deterministically from $X$ and $Y$, allowing us to use the same trick. Labelling outcomes we have:

\begin{center}
\begin{tabular}{|c|c|c|}
\hline
$X$ & $Y$ & $\omega$ \\
\hline
0 & 0 & 1 \\
0 & 1 & 2 \\
1 & 0 & 3 \\
1 & 1 & 4 \\
2 & 0 & 5 \\
2 & 1 & 6 \\
\hline
\end{tabular}
\end{center}

In this case we see that

\begin{equation}
\Delta X \cap \Delta Y = \langle 14, 16, 23, 25, 36, 45 \rangle.
\end{equation}

Again, this is a mutual information and hence positive. Thus we know that for $Z$ to be purely negative it must be generated by purely degree 3 atoms so $Z$ must remain unchanged on these pairs of outcomes. Assigning a symbol to $Z(0,0)$, we can use the same trick as before and successively annihilate various pairs of atoms, giving us the chain:
\newpage

\begin{align}
& Z(0,0) = 0 \notag \\
\implies & Z(1,1) = 0, \quad Z(2,1) = 0 \notag \\
\implies & Z(2,0) = 0,\quad  Z(1,0) = 0 \notag \\
\implies & Z(0,1) = 0
\end{align}

That is to say, $Z$ does not vary on $\Omega$ and the co-information is zero in this case.

We now suppose for the induction that there are no fixed-parity systems with $N_X$ outcomes on $X$ and $N_Y$ outcomes on $Y$. We will demonstrate that we can introduce an additional event to either $X$ or $Y$ and we will still obtain that $Z$ is the trivial variable.

Without loss of generality we increase $N_X$ by one. This will introduce $N_Y$ additional outcomes. As we have (without reference to the probabilities) demonstrated that $Z(\omega_1) = 0$ for $\omega_1 \in S_1 = \{1,\ldots, N_XN_Y\}$, it suffices to show that for every further outcome $\omega_2 \in S_2 = \{N_XN_Y+1,\ldots, (N_X+1)N_Y\}$ there is an atom $\omega_1\omega_2$ with $\omega_1 \in S_1$.

For each $\omega_2$ we shall pick some $\omega_1 \in S_1$ such that $\omega_1\omega_2 \in \Delta X \cap \Delta Y$. Using the ordering we have utilised so far and restricting to the bottom of the table, we may assume that $X(\omega_2) = N_X$ as a symbol, and $Y(\omega_1) \in \{0,\ldots, N_Y-1\}$.

Given $\omega_2$ we may select the outcome $\omega_1$ to be the outcome corresponding to $X = X(\omega_2) - 1$ and $Y = Y(\omega_2) - 1$, where in $Y$ we perform arithmetic mod $N_Y$. We must then have that $X$ and $Y$ both change when moving from $\omega_1$ to $\omega_2$. By the definition of content, this means that the $\omega_1\omega_2$ atom will be contained in $\Delta X \cap \Delta Y$, as needed, so we must have $Z(\omega_2) = Z(\omega_1) = 0$, showing that $Z$ is actually the trivial variable, inductively giving the result.
\end{proof}

\end{appendices}
\end{document}